\documentclass[12pt]{article}
\usepackage{epsfig}
\usepackage{amsmath}
\usepackage{hhline}
\usepackage{amssymb}
\usepackage{times}
\usepackage{cite}

\newlength{\dinwidth}
\newlength{\dinmargin}
\begin{document}  
\newcommand{\pom}{{I\!\!P}}
\newcommand{\reg}{{I\!\!R}}
\newcommand{\slowpi}{\pi_{\mathit{slow}}}
\newcommand{\fiidiii}{F_2^{D(3)}}
\newcommand{\fiidiiiarg}{\fiidiii\,(\beta,\,Q^2,\,x)}
\newcommand{\n}{1.19\pm 0.06 (stat.) \pm0.07 (syst.)}
\newcommand{\nz}{1.30\pm 0.08 (stat.)^{+0.08}_{-0.14} (syst.)}
\newcommand{\fiidiiiful}{F_2^{D(4)}\,(\beta,\,Q^2,\,x,\,t)}
\newcommand{\fiipom}{\tilde F_2^D}
\newcommand{\ALPHA}{1.10\pm0.03 (stat.) \pm0.04 (syst.)}
\newcommand{\ALPHAZ}{1.15\pm0.04 (stat.)^{+0.04}_{-0.07} (syst.)}
\newcommand{\fiipomarg}{\fiipom\,(\beta,\,Q^2)}
\newcommand{\pomflux}{f_{\pom / p}}
\newcommand{\nxpom}{1.19\pm 0.06 (stat.) \pm0.07 (syst.)}
\newcommand {\gapprox}
   {\raisebox{-0.7ex}{$\stackrel {\textstyle>}{\sim}$}}
\newcommand {\lapprox}
   {\raisebox{-0.7ex}{$\stackrel {\textstyle<}{\sim}$}}
\def\gsim{\,\lower.25ex\hbox{$\scriptstyle\sim$}\kern-1.30ex%
\raise 0.55ex\hbox{$\scriptstyle >$}\,}
\def\lsim{\,\lower.25ex\hbox{$\scriptstyle\sim$}\kern-1.30ex%
\raise 0.55ex\hbox{$\scriptstyle <$}\,}
\newcommand{\pomfluxarg}{f_{\pom / p}\,(x_\pom)}
\newcommand{\dsf}{\mbox{$F_2^{D(3)}$}}
\newcommand{\dsfva}{\mbox{$F_2^{D(3)}(\beta,Q^2,x_{I\!\!P})$}}
\newcommand{\dsfvb}{\mbox{$F_2^{D(3)}(\beta,Q^2,x)$}}
\newcommand{\dsfpom}{$F_2^{I\!\!P}$}
\newcommand{\gap}{\stackrel{>}{\sim}}
\newcommand{\lap}{\stackrel{<}{\sim}}
\newcommand{\fem}{$F_2^{em}$}
\newcommand{\tsnmp}{$\tilde{\sigma}_{NC}(e^{\mp})$}
\newcommand{\tsnm}{$\tilde{\sigma}_{NC}(e^-)$}
\newcommand{\tsnp}{$\tilde{\sigma}_{NC}(e^+)$}
\newcommand{\st}{$\star$}
\newcommand{\sst}{$\star \star$}
\newcommand{\ssst}{$\star \star \star$}
\newcommand{\sssst}{$\star \star \star \star$}
\newcommand{\tw}{\theta_W}
\newcommand{\sw}{\sin{\theta_W}}
\newcommand{\cw}{\cos{\theta_W}}
\newcommand{\sww}{\sin^2{\theta_W}}
\newcommand{\cww}{\cos^2{\theta_W}}
\newcommand{\trm}{m_{\perp}}
\newcommand{\trp}{p_{\perp}}
\newcommand{\trmm}{m_{\perp}^2}
\newcommand{\trpp}{p_{\perp}^2}
\newcommand{\alp}{\alpha_s}

\newcommand{\alps}{\alpha_s}
\newcommand{\sqrts}{$\sqrt{s}$}
\newcommand{\LO}{$O(\alpha_s^0)$}
\newcommand{\Oa}{$O(\alpha_s)$}
\newcommand{\Oaa}{$O(\alpha_s^2)$}
\newcommand{\PT}{p_{\perp}}
\newcommand{\JPSI}{J/\psi}
\newcommand{\sh}{\hat{s}}
\newcommand{\uh}{\hat{u}}
\newcommand{\MP}{m_{J/\psi}}
\newcommand{\PO}{I\!\!P}
\newcommand{\xbj}{x}
\newcommand{\xpom}{x_{\PO}}
\newcommand{\ttbs}{\char'134}
\newcommand{\xpomlo}{3\times10^{-4}}  
\newcommand{\xpomup}{0.05}  
\newcommand{\dgr}{^\circ}
\newcommand{\pbarnt}{\,\mbox{{\rm pb$^{-1}$}}}
\newcommand{\gev}{\,\mbox{GeV}}
\newcommand{\WBoson}{\mbox{$W$}}
\newcommand{\fbarn}{\,\mbox{{\rm fb}}}
\newcommand{\fbarnt}{\,\mbox{{\rm fb$^{-1}$}}}
\newcommand{\dsdx}[1]{$d\sigma\!/\!d #1\,$}
\newcommand{\eV}{\mbox{e\hspace{-0.08em}V}}
%
%
\newcommand{\qsq}{\ensuremath{Q^2} }
\newcommand{\gevsq}{\ensuremath{\mathrm{GeV}^2} }
\newcommand{\et}{\ensuremath{E_t^*} }
\newcommand{\rap}{\ensuremath{\eta^*} }
\newcommand{\gp}{\ensuremath{\gamma^*}p }
\newcommand{\dsiget}{\ensuremath{{\rm d}\sigma_{ep}/{\rm d}E_t^*} }
\newcommand{\dsigrap}{\ensuremath{{\rm d}\sigma_{ep}/{\rm d}\eta^*} }

\newcommand{\dstar}{\ensuremath{D^*}}
\newcommand{\dstarp}{\ensuremath{D^{*+}}}
\newcommand{\dstarm}{\ensuremath{D^{*-}}}
\newcommand{\dstarpm}{\ensuremath{D^{*\pm}}}
\newcommand{\zDs}{\ensuremath{z(\dstar )}}
\newcommand{\Wgp}{\ensuremath{W_{\gamma p}}}
\newcommand{\ptds}{\ensuremath{p_t(\dstar )}}
\newcommand{\etads}{\ensuremath{\eta(\dstar )}}
\newcommand{\ptj}{\ensuremath{p_t(\mbox{jet})}}
\newcommand{\ptjn}[1]{\ensuremath{p_t(\mbox{jet$_{#1}$})}}
\newcommand{\etaj}{\ensuremath{\eta(\mbox{jet})}}
\newcommand{\detadsj}{\ensuremath{\eta(\dstar )\, \mbox{-}\, \etaj}}

\def\Journal#1#2#3#4{{#1} {\bf #2} (#3) #4}
\def\NCA{\em Nuovo Cimento}
\def\NIM{\em Nucl. Instrum. Methods}
\def\NIMA{{\em Nucl. Instrum. Methods} {\bf A}}
\def\NPB{{\em Nucl. Phys.}   {\bf B}}
\def\PLB{{\em Phys. Lett.}   {\bf B}}
\def\PRL{\em Phys. Rev. Lett.}
\def\PRD{{\em Phys. Rev.}    {\bf D}}
\def\ZPC{{\em Z. Phys.}      {\bf C}}
\def\EJC{{\em Eur. Phys. J.} {\bf C}}
\def\CPC{\em Comp. Phys. Commun.}

\newcommand{\be}{\begin{equation}} 
\newcommand{\ee}{\end{equation}} 
\newcommand{\ba}{\begin{eqnarray}} 
\newcommand{\ea}{\end{eqnarray}}

\begin{titlepage}

\noindent
\begin{flushleft}
{\tt DESY 08-009    \hfill    ISSN 0418-9833} \\
{\tt January 2008}                  \\
\end{flushleft}

\vspace{2cm}
\begin{center}
\begin{Large}

{\bf A Search for Excited Neutrinos in \begin{boldmath}$e^{-}p$\end{boldmath}~ Collisions at HERA \\}

\vspace{2cm}

H1 Collaboration

\end{Large}
\end{center}

\vspace{2cm}

\begin{abstract}
A search for excited neutrinos is performed using the full  $e^{-}p$~data sample collected by the H1 experiment at HERA at a centre-of-mass energy of $319$ GeV, corresponding to a total luminosity of $184$~pb$^{-1}$.
The electroweak decays of excited neutrinos ${\nu}^{*}{\rightarrow}{\nu}{\gamma}$, ${\nu}^{*}{\rightarrow}{\nu}Z$ and ${\nu}^{*}{\rightarrow}eW$ with subsequent hadronic or leptonic decays of the $W$ and $Z$ bosons are considered.
No evidence for excited neutrino production is found. 
Mass dependent exclusion limits on $\nu^*$ production cross sections and on the ratio of the coupling to the compositeness scale $f/{\Lambda}$ are derived within gauge mediated models. 
A limit on $f/{\Lambda}$, independent of the relative couplings to the SU($2$) and U($1$) gauge bosons, is also determined.
These limits extend the excluded region to higher masses than has been possible in previous excited neutrino searches.
\end{abstract}

\vspace{1.5cm}

\begin{center}
Submitted to \PLB
\end{center}

\end{titlepage}

%
%
%
\begin{flushleft}

F.D.~Aaron$^{5,49}$,           
C.~Alexa$^{5}$,                
V.~Andreev$^{25}$,             
B.~Antunovic$^{11}$,           
S.~Aplin$^{11}$,               
A.~Asmone$^{33}$,              
A.~Astvatsatourov$^{4}$,       
S.~Backovic$^{30}$,            
A.~Baghdasaryan$^{38}$,        
P.~Baranov$^{25, \dagger}$
% ...

\bigskip{\it
 $ ^{1}$ I. Physikalisches Institut der RWTH, Aachen, Germany$^{ a}$ \\
% ...
$ ^{4}$ Inter-University Institute for High Energies ULB-VUB, Brussels;
          Universiteit Antwerpen, Antwerpen; Belgium$^{ c}$ \\
 $ ^{5}$ National Institute for Physics and Nuclear Engineering (NIPNE) ,
          Bucharest, Romania \\
% ...
$ ^{11}$ DESY, Hamburg, Germany \\
% ...
$ ^{25}$ Lebedev Physical Institute, Moscow, Russia$^{ e}$ \\
% ...
$ ^{30}$ Faculty of Science, University of Montenegro,
          Podgorica, Montenegro$^{ e}$ \\
% ...
$ ^{33}$ Dipartimento di Fisica Universit\`a di Roma Tre
          and INFN Roma~3, Roma, Italy \\
% ...
$ ^{38}$ Yerevan Physics Institute, Yerevan, Armenia \\
% ...
$ ^{49}$ Also at Faculty of Physics, University of Bucharest,
          Bucharest, Romania \\
% ...
$ ^{\dagger}$ Deceased \\

\bigskip
 $ ^a$ Supported by the Bundesministerium f\"ur Bildung und Forschung, FRG,
      under contract numbers 05 H1 1GUA /1, 05 H1 1PAA /1, 05 H1 1PAB /9,
      05 H1 1PEA /6, 05 H1 1VHA /7 and 05 H1 1VHB /5 \\
% ...
$ ^c$ Supported by FNRS-FWO-Vlaanderen, IISN-IIKW and IWT
      and  by Interuniversity
Attraction Poles Programme,
      Belgian Science Policy \\
% ...
$ ^e$ Supported by the Deutsche Forschungsgemeinschaft \\
% ...
}

\end{flushleft}
%

\newpage

\section{Introduction}

The three-family structure and mass hierarchy of the known fermions is one of the most puzzling characteristics of the Standard Model (SM) of particle physics.
Attractive explanations are provided by models assuming composite quarks and leptons~\cite{Harari:1982xy}.
The existence of excited states of leptons and quarks is a natural consequence of these models and their discovery would provide convincing evidence of a new scale of matter.
Electron-proton interactions at very high energies provide a good environment in which  to search for excited states of first generation fermions.
In particular, excited neutrinos ($\nu^*$) could be singly produced through the exchange of a $W$ boson in the $t$-channel. 

In this letter a search for excited neutrinos using the complete $e^{-}p$ HERA collider data of the H1 experiment is presented. 
Electroweak decays into a SM lepton ($e$, $\nu_e$) and a SM gauge boson ($\gamma$, $W$ and $Z$) are considered and hadronic as well as leptonic decays of the $W$ and $Z$ are analysed.

The data, collected at electron and proton beam energies of $27.6$~GeV and $920$~GeV, respectively, correspond to a total integrated luminosity of $184$~pb$^{-1}$. 
During the HERA~II running period, the electron beam was longitudinally polarised, at a level of typically $35\%$. For this analysis the periods with left-handed and right-handed beams are combined and the analysed data sample has a residual polarisation of $5\%$ left-handed.
With more than a ten-fold increase in statistics, this analysis supersedes the result of the previous H1 search for excited neutrinos based on a data sample corresponding to a luminosity of $15$~pb$^{-1}$~\cite{Adloff:2001me}.

\section{Phenomenology}

In the present study a model~\cite{Hagiwara:1985wt,Boudjema:1992em,Baur:1989kv} is considered in which excited fermions are assumed to have spin  $1/2$ and isospin $1/2$.
Both left-handed and right-handed components of the excited fermions form weak iso-doublets $F_L^*$ and $F_R^*$.
In order to prevent the light leptons from radiatively acquiring a large anomalous magnetic moment~\cite{Brodsky:1980zm,Renard:1982ij}, only the right-handed component of the excited fermions takes part in the generalised magnetic de-excitation. 
The interaction between excited fermions, gauge bosons and ordinary fermions is then described by the effective Lagrangian~\cite{Boudjema:1992em}:

\be
{\cal L}_{int.} = \frac{1}{2\Lambda}{\bar{F^{*}_{R}}} \; {{\sigma}^{\mu\nu}} \left[ gf\frac{\tau^a}{2}{W_{\mu\nu}^a}+g'f'\frac{Y}{2}B_{\mu\nu} + g_s f_s \frac{\lambda^a}{2} G^a_{\mu\nu} \right] \; {F_{L}} + h.c. \; .
\label{eq:lagrangian}
\ee

The matrix ${{\sigma}^{\mu\nu}}$ is the covariant bilinear tensor, $W_{{\mu}{\nu}}^a$,  $B_{{\mu}{\nu}}$ and $G^a_{\mu\nu}$ are the field-strength tensors of the SU($2$), U($1$) and SU($3$)$_{C}$ gauge fields, $\tau^a$, $Y$ and $\lambda^a$ are the Pauli matrices, the weak hypercharge operator and the Gell-Mann matrices, respectively. The standard electroweak and strong gauge couplings are denoted by $g$, $g'$ and $g_s$, respectively.
The parameter $\Lambda$ has units of energy and can be regarded as the compositeness scale which reflects the range of a new confinement force. The constants $f$, $f'$ and $f_s$ are form factors associated to the three gauge groups. 
They can be interpreted as parameters setting different scales $\Lambda_i = \Lambda/f_i$ for the different gauge groups, thus allowing the composite fermion to have arbitrary coupling strengths with the three gauge bosons.

Following this model, single production of excited neutrinos in $ep$ collisions may result from the $t$-channel exchange of a $W$ boson. 
Due to the helicity dependence of the weak interaction and given the valence quark composition and density distribution of the proton, the $\nu^*$ production cross section is predicted to be much larger for $e^-p$ collisions than for  $e^+p$. For a $\nu^*$ mass $M_{\nu^*}$ of $200$~GeV the ratio of the production cross sections is of order $100$.
The $\nu^*$ production cross section is expected to scale linearly with the polarisation of the incident electron beam, similarly to the SM charged current process.
The excited neutrino may decay into a lepton and an electroweak gauge boson via $\nu^*$~${\rightarrow} \nu\gamma$, $\nu^* {\rightarrow} eW$ and $\nu^* {\rightarrow} \nu Z$.
For a given $M_{\nu^*}$ value and assuming a numerical relation between $f$ and $f'$, the $\nu^*$ branching ratios are fixed and the production cross section depends only on $f/\Lambda$.
%
The $\nu^*$ is expected not to have strong interactions and therefore this search is insensitive to $f_s$.
%
Two complementary coupling assignments $f = + f'$ and $f = - f'$ are studied in detail. 
For $f = + f'$ the excited neutrino has no tree-level electromagnetic coupling~\cite{Boudjema:1989yx} and therefore the photonic decay of the $\nu^*$ is forbidden whereas for $f = - f'$ decays into $\nu\gamma$, $\nu{Z}$ and $eW$ are allowed.
In addition, arbitrary ratios of $f'/f$ are considered in the range $-5$ to $+5$.

\section{Simulation of Signal and Background Processes}

A Monte Carlo (MC) program developed for this analysis is used for the calculation of the $\nu^*$ production cross section and the simulation of signal events. 
The events are simulated using the cross section calculated from the Lagrangian described in equation (\ref{eq:lagrangian}) using the CompHEP package~\cite{comphep}. 
Initial state radiation from the incident electron is included using the Weizs\"acker-Williams approximation~\cite{Berger:1986ii}. 
The proton parton densities are taken from the CTEQ5L~\cite{Pumplin:2002vw} parametrisation and are evaluated at the scale $\sqrt{Q^2}$, where $Q^2$ is the four-momentum transfer squared. The parton shower approach~\cite{Sjostrand:2000wi} is applied to simulate Quantum Chromodynamics (QCD) corrections in the initial and final states. The hadronisation is performed using Lund string fragmentation as implemented in PYTHIA~\cite{Sjostrand:2000wi}.
In the MC generator the full transition matrix including the production and the decay is implemented. 
This is important if the natural width of the $\nu^*$ is large, which is typically the case at high mass where factorisation of the $\nu^*$ production and its decay no longer holds. 
Events used in the determination of signal efficiencies are generated with the coupling $f/\Lambda$ corresponding, for each $\nu^*$ mass, to the expected boundary of the probed domain in the plane 
defined by $M_{\nu^*}$ and  $f/\Lambda$.

%
%
The Standard Model background processes that may mimic a $\nu^*$ signal are neutral current (NC) and charged current (CC) deep-inelastic scattering (DIS) and to a lesser extent photoproduction, lepton pair production and real $W$ boson production. 

The RAPGAP~\cite{Jung:1993gf} event generator, which implements the Born, QCD Compton and Boson Gluon Fusion matrix elements, is used to model NC DIS events. 
The QED radiative effects arising from real photon emission from both the incoming and outgoing electrons are simulated using the HERACLES~\cite{Kwiatkowski:1990es} program. 
Direct and resolved photoproduction of jets and prompt photon production are simulated using the PYTHIA event generator. 
The simulation is based on Born level hard scattering matrix elements with radiative QED corrections. 
In RAPGAP and PYTHIA, jet production from higher order QCD radiation is simulated using leading logarithmic parton showers and hadronisation is modelled with Lund string fragmentation.
The leading order MC prediction of NC DIS and photoproduction processes with two or more high transverse momentum jets is scaled by a factor of $1.2$ in order to normalise to next-to-leading order QCD calculations~\cite{Adloff:2002au}. 
Charged current DIS events are simulated using the DJANGO~\cite{Schuler:yg} program, which includes first order leptonic QED radiative corrections based on HERACLES. The production of two or more jets in DJANGO is accounted for using the colour-dipole-model~\cite{Lonnblad:1992tz}. 

Contributions from elastic and quasi-elastic QED Compton scattering are simulated with the WABGEN~\cite{Berger:kp} generator. 
Contributions arising from the production of $W$ bosons and multi-lepton events are modelled using the EPVEC~\cite{Baur:1991pp} and GRAPE~\cite{Abe:2000cv} event generators, respectively.

All processes are generated with an integrated luminosity significantly higher than that of the data sample.
Generated events are passed through the full GEANT~\cite{Brun:1987ma} based simulation of the H1 apparatus, which takes into account the running conditions of the different data taking periods, and are reconstructed and analysed using the same program chain as for the data.

\section{Experimental Conditions}

A detailed description of the H1 experiment can be found in~\cite{Abt:h1}.
Only the detector components relevant to the
present analysis are briefly described here.  
The origin of the H1 coordinate system is the nominal $ep$ interaction point, with the direction of the proton beam defining the positive $z$-axis (forward region). Transverse momentum ($P_T$) is measured in the $xy$ plane. The pseudorapidity $\eta$ is related to the polar angle $\theta$ by $\eta = -\ln \, \tan (\theta/2)$.
The Liquid Argon (LAr) calorimeter~\cite{Andrieu:1993kh} is used to measure electrons, photons and hadrons. It covers the polar angle range
$4^\circ < \theta < 154^\circ$ with full azimuthal acceptance.
Electromagnetic shower energies are measured with a precision of
$\sigma (E)/E = 12\%/ \sqrt{E/\mbox{GeV}} \oplus 1\%$ and hadronic energies
with $\sigma (E)/E = 50\%/\sqrt{E/\mbox{GeV}} \oplus 2\%$, as measured in test beams~\cite{Andrieu:1993tz,Andrieu:1994yn}.
In the backward region, energy measurements are provided by a lead/scintillating-fiber (SpaCal) calorimeter~\cite{Appuhn:1996na} covering the angular range $155^\circ < \theta < 178^\circ$.
The central ($20^\circ < \theta < 160^\circ$) and forward ($7^\circ < \theta < 25^\circ$)  tracking detectors are used to
measure charged particle trajectories, to reconstruct the interaction
vertex and to complement the measurement of hadronic energy.
The LAr and inner tracking detectors are enclosed in a super-conducting magnetic
coil with a field strength of $1.16$~T.
The return yoke of the coil is the outermost part of the detector and is
equipped with streamer tubes forming the central muon detector
($4^\circ < \theta < 171^\circ$).
In the forward region of the detector ($3^\circ < \theta < 17^\circ$) a set of
drift chambers detects muons and measures their momenta using an iron toroidal magnet.
The luminosity is determined from the rate of the Bethe-Heitler process $ep {\rightarrow} ep \gamma$,
measured using a photon detector located close to the beam pipe at $z=-103~{\rm m}$, in the backward direction.

\section{Data Analysis}

The triggers used in this analysis are based on the detection of energy deposits in the LAr calorimeter~\cite{Adloff:2003uh}.
Events containing an electromagnetic deposit (electron or photon) with an energy greater than $10$~GeV are triggered with an efficiency close to $100$\%.
For events with missing transverse energy above $20$~GeV, the trigger efficiency is $\sim$~$90$\%.

In order to remove background events induced by cosmic showers and other non-$ep$ sources, the event vertex is required to be reconstructed within $35$~cm in $z$ of the nominal interaction point. In addition, topological filters and timing vetoes are applied.

The identification of electrons or photons relies on the measurement of a compact and isolated electromagnetic shower in the LAr calorimeter. 
In addition, the hadronic energy within a distance in the pseudorapidity-azimuth $(\eta - \phi)$ plane $R=\sqrt{\Delta \eta^2 + \Delta \phi^2} < 0.5$ around the electron (photon) is required to be below $3$\% of the electron (photon) energy.
Muon identification is based on a track measured in the inner tracking systems associated with signals in the muon detectors~\cite{Andreev:2003pm}.
A muon candidate should have no more than $5$~GeV deposited in a
cylinder, centred on the muon track direction, of radius $25$~cm and $50$~cm in the electromagnetic and hadronic sections of the LAr calorimeter, respectively.
Calorimeter energy deposits and tracks not previously identified as electron, photon or muon candidates are used to form combined cluster-track objects, from which the hadronic energy is reconstructed~\cite{matti,benji}.
Jets are reconstructed from these combined cluster-track objects using an inclusive $k_T$ algorithm~\cite{Ellis:1993tq,Catani:1993hr} with a minimum transverse momentum of $2.5$~GeV.
The missing transverse momentum  $P_T^{\rm{miss}}$ of the event is derived from all identified particles and energy deposits in the event.
%
The $P_T^{\rm{miss}}$ is assumed to originate from a single neutrino.
The four-vector of this neutrino candidate is reconstructed assuming transverse momentum conservation and the relation $\sum_i (E^i - P_{z}^{i}) + (E^\nu - P_{z}^{\nu}) = 2 E^0_e = 55.2$~GeV, where the sum runs over all detected particles; $P_{z}$ is the momentum along the beam axis and $E^0_e$ is the electron beam energy.

Specific selection criteria applied in each decay channel are presented in the following subsections.
A detailed description of the analysis can be found in~\cite{trinh}.

\subsection{$\nu\gamma$ Resonance Search}

The signature of the decay channel $\nu^* {\rightarrow} \nu \gamma$ consists of an isolated electromagnetic cluster in events with missing transverse momentum. 
Background arises from CC DIS events with an isolated ${\pi}^0$ or a radiated photon.
Events with substantial missing transverse momentum \mbox{$P_{T}^{\rm{miss}} > 20$~GeV} are selected.
In each event, a photon candidate with transverse momentum \mbox{$P_T^\gamma >$ $20$~GeV} in a polar angle range $5^{\circ} < {\theta}^{\gamma} < 120^{\circ}$ is required. 
This polar angle range is restricted to ${\theta}^{\gamma} < 60^{\circ}$ in events with $P_T^{\rm{miss}}$ below $30$~GeV, in order to reduce background from NC DIS.
The photon is required to be isolated from jets by a distance $R > 0.5$ to any jet axis.
In the central region (${\theta}^{\gamma} > 20^{\circ}$), photon candidates are selected only if no well measured track points to the electromagnetic cluster within a distance of closest approach (DCA) of $12$~cm.
For events with $P_{T}^{\rm{miss}}$ below $50$~GeV, this condition is tightened by accepting only photon candidates having no track with a DCA to the cluster below $24$~cm or within $R <0.5$.
The energy and polar angle of the photon are combined into one discriminant variable $\xi^\gamma = E^\gamma \cos^2{(\theta^\gamma/2)}$.
Radiative CC DIS events are suppressed by requiring that  $\xi^\gamma > 45$~GeV. %
For signal events, in most cases the final state contains a recoil jet, due to ${\nu}^{*}$ production via $t$-channel $W$~boson exchange. 
Hence, in the final selection the presence of at least one jet with $P_{T}^{\rm{jet}} > 5$~GeV is also required.

\begin{figure}[!htbp] 
   \begin{center}
\includegraphics[width=.5\textwidth]{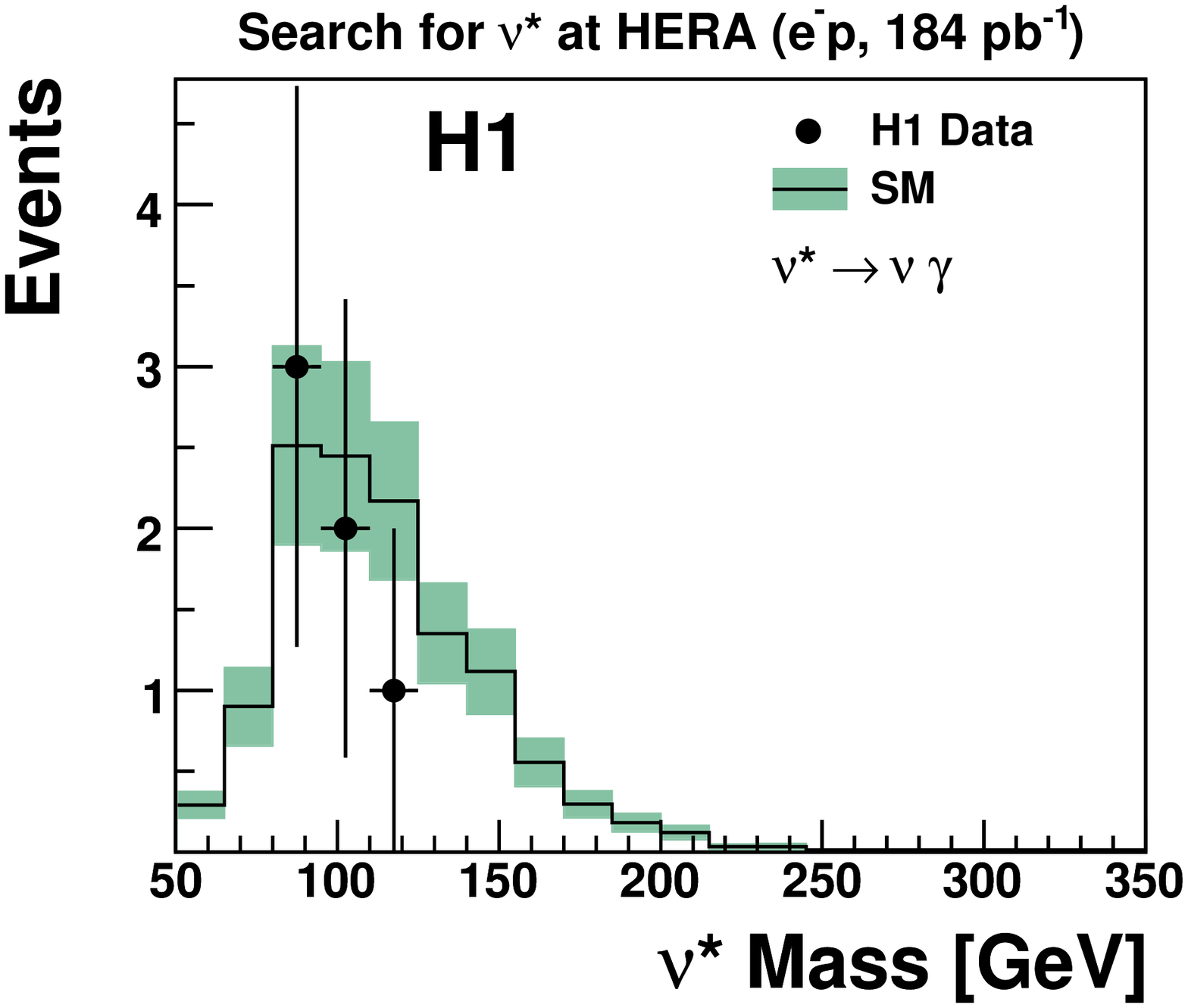}\put(-12,37) {{\bf (a)}}
\includegraphics[width=.5\textwidth]{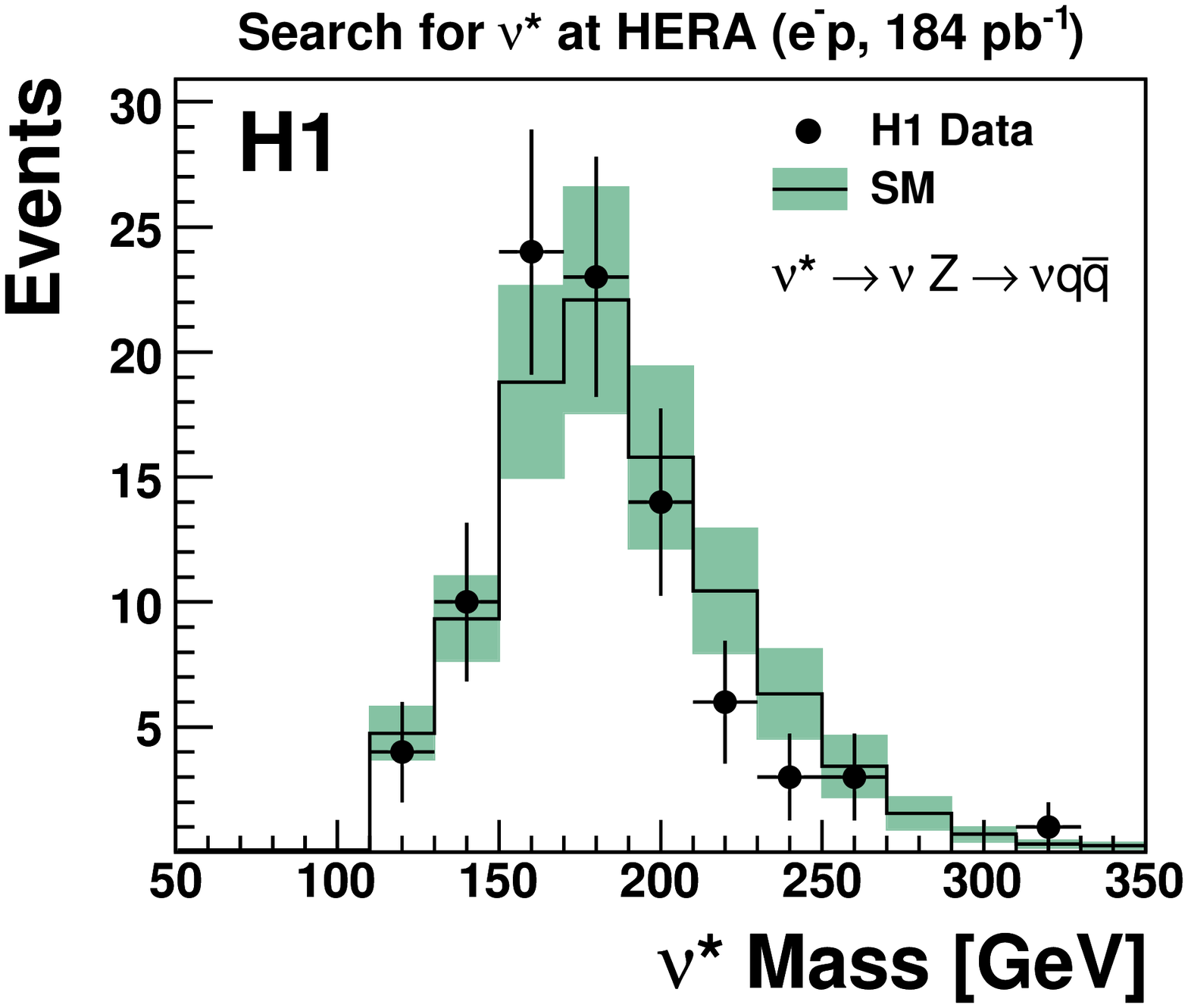}\put(-12,37) {{\bf (b)}}\\ 
\hspace{-.5\textwidth}\includegraphics[width=.5\textwidth]{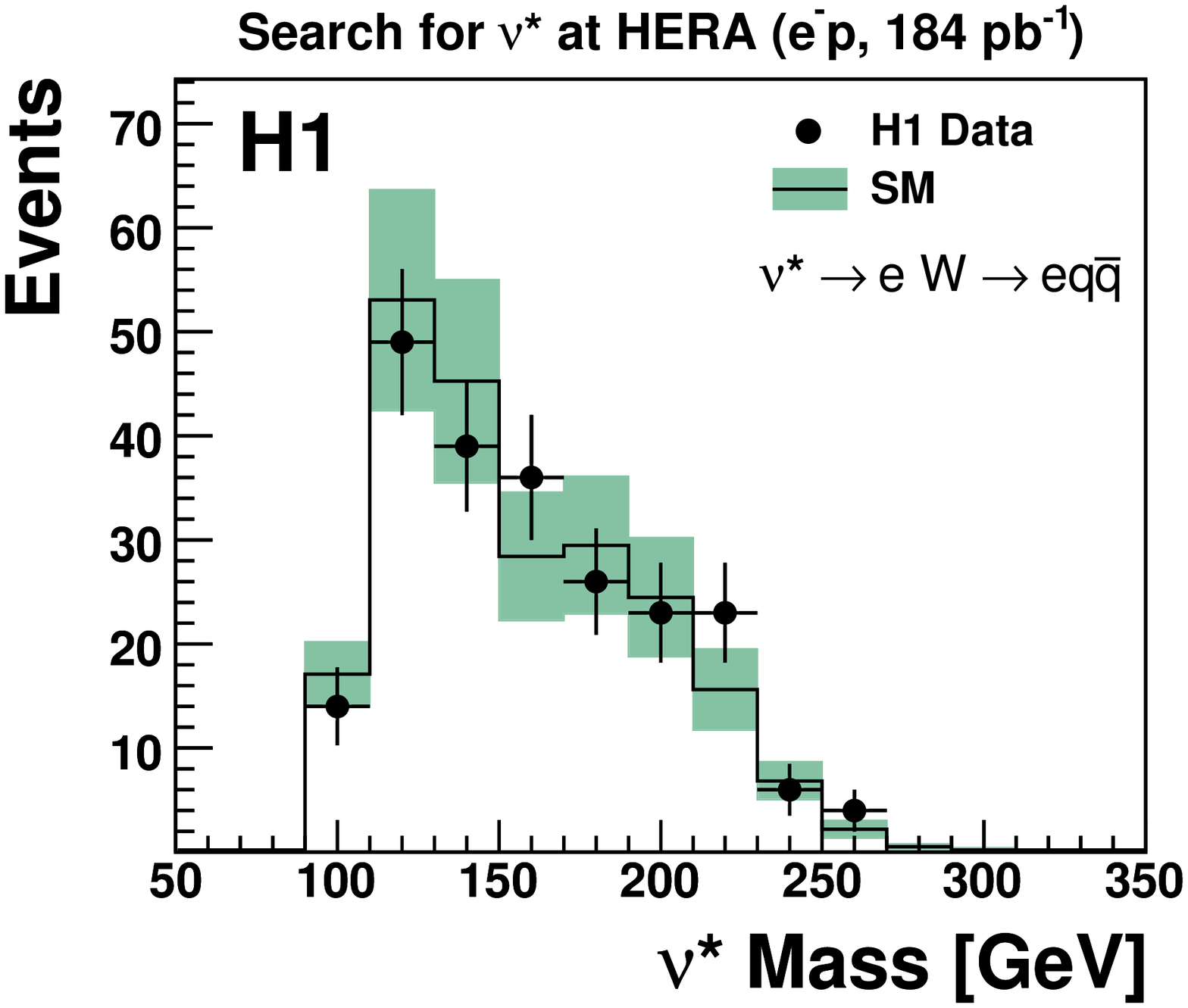}\put(-12,37) {{\bf (c)}}
\end{center}
 \caption{Invariant mass distribution of the $\nu^*$ candidates for the ${\nu}^{*} {\rightarrow} {\nu}{\gamma}$ (a), ${\nu}^{*} {\rightarrow} {\nu} Z {\rightarrow} \nu q\bar{q}$ (b) and ${\nu}^{*} {\rightarrow} {e} W {\rightarrow} e q\bar{q}$ (c) searches. The points correspond to the observed data events and the histogram to the SM expectation after the final selections. The error bands on the SM prediction include model and experimental systematic errors added in quadrature.}
 \label{fig:Mass}  
 \end{figure}

Seven events are selected in the data, compared to a SM expectation of $12.3$~$\pm$~$3.0$, which is dominated by CC DIS events.
The invariant mass of the excited neutrino candidate is calculated from the four-vectors of the neutrino and the photon. 
The invariant mass distributions of the $\nu^*$ candidates and of the expected SM background are presented in figure~\ref{fig:Mass}(a).
The resulting selection efficiency is $50$\% for $M_{\nu^*} = 120$~GeV, increasing to $55$\% for $M_{\nu^*} = 260$~GeV.
From Monte Carlo studies, the total width of the reconstructed $\nu^*$ mass distribution is $11$~GeV for a generated $\nu^*$ mass of $120$~GeV, increasing to $41$~GeV for a $\nu^*$ mass of $260$~GeV. 

\subsection{$\nu{q}{\bar{q}}$  Resonance Search}

The signature of the ${\nu}^{*} {\rightarrow} \nu Z {\rightarrow} \nu q \bar{q}$ decay channel consists of two jets with high transverse momentum in events with large $P_T^{\rm{miss}}$.
The SM background is dominated by multi-jet CC DIS events and contains a moderate contribution from photoproduction.
Events with missing transverse momentum $P_{T}^{\rm{miss}} > 20$~GeV are selected.
In each event at least two jets are required in the polar angle range $5^{\circ} < \theta^{\rm{jet}} < 130^{\circ}$ with transverse momenta larger than $20$ and $15$~GeV, respectively.
Additionally, the hadronic final state  must exhibit a polar angle $\gamma_h$, as defined in~\cite{Adloff:1999ah}, larger than $20^{\circ}$, in order to remove photoproduction events.
Events with  
$P_{T}^{\rm{miss}} <$~$30$~GeV are selected if the ratio $V_{ap}/V_{p}$ of transverse energy flow anti-parallel and parallel to the hadronic final state~\cite{Adloff:1999ah} is above $0.1$.
This condition reduces the contribution of CC DIS processes. 
Photoproduction and NC DIS backgrounds typically have low volues of $x_h$, the Bjorken scaling variable calculated from the hadronic system using the Jacquet-Blondel method~\cite{Adloff:1999ah,JBmethod}, and are thus suppressed by requiring $x_h > 0.04$.
Finally, to further reduce the background from CC DIS, a jet multiplicity greater than or equal to three is required for events with $P_{T}^{\rm{miss}} < 50$~GeV.  
In each event, a $Z$ candidate is reconstructed from the combination of the two jets with an invariant mass closest to the nominal $Z$ boson mass. The reconstructed $Z$ candidate is required to have an invariant mass above $60$~GeV.

After this selection, 
$89$ events are found in the data compared to a SM expectation of $95$~$\pm$~$21$ events. 
The invariant mass of the $\nu^*$ candidate is calculated from the neutrino and $Z$ candidate four-vectors.
For this calculation, the $Z$ candidate four-vector is scaled such that its mass is set equal to the nominal $Z$ boson mass.
The invariant mass distributions of the $\nu^*$ candidates and of the expected SM background are presented in figure~\ref{fig:Mass}(b).
The selection efficiency in this channel is $25$\% for $M_{\nu^*} = 120$~GeV, increasing to $55$\% for $M_{\nu^*} = 260$~GeV.
From Monte Carlo studies, the total width of the reconstructed $\nu^*$ mass distribution is $31$~GeV for a generated $\nu^*$ mass of $120$~GeV, increasing to $41$~GeV for a $\nu^*$ mass of $260$~GeV.

\subsection{$eq{\bar{q}}$ Resonance Search} 

The signature of the ${\nu}^{*} {\rightarrow} e W {\rightarrow} e q \bar{q}$ decay channel consists of one electron and two high $P_T$ jets.
Multi-jet NC DIS events constitute the main background contribution from SM processes.
Events are selected with an isolated electron in the LAr calorimeter in the polar angle range $5^\circ< \theta^e < 90^\circ$. 
The electron variable\footnote{ This variable is proportional to the four-momentum transfer squared $Q^2$ for NC DIS.} $\xi^e = E^e \cos^2{(\theta^e/2)}$ is required to be above $23$~GeV or the electron should have a transverse momentum $P_T^e$ greater than $25$~GeV.
These conditions remove a large part of the NC DIS contribution.
In addition, the electron should be isolated from jets by a distance $R > 0.5$ to any jet axis.
The events are required to have at least two jets in the polar angle range $5^{\circ} < \theta^{\rm{jet}} < 130^{\circ}$ with transverse momenta larger than $20$ and $15$~GeV, respectively.
In each event, a $W$ candidate is reconstructed from the combination of the two jets with an invariant mass closest to the nominal $W$ boson mass. 
The reconstructed mass of the $W$ candidate is required to be larger than $40$~GeV.
To further reduce the NC DIS background it is required that the polar angle of the jet with the highest $P_T$ associated to the $W$ candidate be less than $80^\circ$ and that events with $P_T^e <$~$65$~GeV contain at least three jets with a $P_T$ larger than $5$~GeV.

After the selection, $220$ events are observed compared to a SM expectation of $223$ $\pm$ $47$.
The invariant mass of the $\nu^*$ candidate is calculated from the electron and $W$ candidate four-vectors. 
For this calculation, the $W$ candidate four-vector is scaled such that its mass is set equal to the nominal $W$ boson mass.
The invariant mass distributions of the $\nu^*$ candidates and of the expected SM background are presented in figure~\ref{fig:Mass}(c).
The selection efficiency in this channel is $40$\% for $M_{\nu^*} = 120$~GeV, increasing to $65$\% for $M_{\nu^*} = 260$~GeV.
From Monte Carlo studies, the total width of the reconstructed $\nu^*$ mass distribution is $15$~GeV for a generated $\nu^*$ mass of $120$~GeV, increasing to $38$~GeV for a $\nu^*$ mass of $260$~GeV.

\subsection{$e \nu \mu$ and $e \nu e$ Resonance Searches} 

In the search for ${\nu}^{*} {\rightarrow} {e} W {\rightarrow} e \nu\mu$, events with $P_T^{\rm{miss}} > 15$~GeV, one electron with $P_T^{e} > 20$~GeV and one muon with $P_T^{\mu} > 10$~GeV are selected. 
The electron and the muon have to be detected in the polar angle ranges $5^\circ < \theta^{e} <  100^\circ$ and $10^\circ< \theta^\mu < 160^\circ$, respectively. 
Furthermore, the electron and the muon must be isolated from jets by minimum distances of  $R^e > 0.5$ and $R^\mu > 1$, respectively.
The contribution from NC DIS processes is reduced by requiring \mbox{$\xi^e >9$~GeV}.
After this selection no data event remains, while $0.40 \pm 0.05$ SM background events are expected. The selection efficiency for $\nu^*$ with masses above $120$~GeV is $\sim 35$\%.

The signatures of the ${\nu}^{*} {\rightarrow} {e} W {\rightarrow} e \nu e$ and ${\nu}^{*} {\rightarrow} {\nu} Z {\rightarrow} \nu e e$ channels are similar and consist of two high $P_T$ electrons in events with large missing transverse momentum.
Events with $P_T^{\rm{miss}} > 20$~GeV are selected. 
In each event two isolated electromagnetic clusters are required, with a transverse momentum larger than $20$ and $15$~GeV, respectively.
The highest $P_T$ electron should be detected in the polar angle range $5^\circ < \theta^{e1} < 100^\circ$ and the second electron in the range $5^\circ < \theta^{e2} < 120^\circ$. 
To reduce the background from Compton processes, a track has to be associated to each electromagnetic cluster in the central region ($\theta^e >35^\circ$).
Events in which the invariant mass of the two electromagnetic clusters is within $10$~GeV of the nominal $Z$ boson mass are attributed to the $Z {\rightarrow} ee$ decay channel. 
Events from the $W {\rightarrow} \nu e$ decay channel are selected by requiring the invariant mass of the $\nu$ and one of the electromagnetic clusters to be compatible with the $W$ boson mass within $20$ GeV.
In this channel, the variable $\xi^e$ defined from the highest $P_T$ electron is required to be above $29$~GeV.
No data candidate is observed in either the $Z$ or $W$ decay channels compared to SM expectations of $0.19 \pm 0.05$ and $0.7 \pm 0.1$, respectively. 
In both channels, the selection efficiency for $\nu^*$ with masses above $120$~GeV is $\sim 35$\%.

\subsection{Systematic Uncertainties}

The following experimental systematic uncertainties are considered:

\begin{itemize}
\item The uncertainty on the electromagnetic energy scale varies between $1$\% and $3$\% depending on the polar angle. The polar angle measurement  uncertainty of electromagnetic clusters is $3$ mrad. 
\item The efficiency to identify photons is known with a precision of $10$\% for high $P_T$ photons.
\item The scale uncertainty on the transverse momentum of high $P_T$ muons amounts to $5$\%. The uncertainty on the reconstruction of the muon polar angle is $3$~mrad.
\item The hadronic energy scale is known within $2$\%. The uncertainty on the jet polar angle determination is $10$~mrad.
\item The uncertainty on the trigger efficiency is $3$\%.
\item The luminosity measurement has an uncertainty of $2.5$\%.
\end{itemize}

The effect of the above systematic uncertainties are determined by varying the experimental quantities by $\pm 1$ standard deviation in the MC samples and propagating these variations through the whole analysis chain.

Additional model systematic uncertainties are attributed to the SM background MC generators described in section $3$.
An error of $20$\% on the simulation of NC DIS, CC DIS and photoproduction processes with at least two high $P_T$ jets is considered to account for the uncertainty on higher order QCD corrections. 
An uncertainty of $30$\% on the simulation of radiative CC DIS events is considered to account for the lack of QED radiation from the quark line in the DJANGO generator. This uncertainty is estimated in the specific phase space of the analysis by a comparison of the DJANGO result to the calculated cross section of the $e^- p {\rightarrow} \nu_e \gamma X$ process~\cite{Helbig:1991iw}.
The error on the QED Compton cross section is estimated to be $10$\%. 
The errors attributed to multi-lepton and $W$ production are $3$\% and $15$\%, respectively.
The total error on the SM background prediction is determined by adding the effects of all model and experimental systematic uncertainties in quadrature.

The theoretical uncertainty on the $\nu^*$ production cross section is dominated by the uncertainty on the scale at which the proton parton densities are evaluated.
It is estimated by varying this scale from $\sqrt{Q^2}/2$ to $2\sqrt{Q^2}$.
The resulting uncertainty depends on the $\nu^*$ mass and is $10$\% at $M_{\nu^*} = 100$~GeV, increasing to $30$\% at $M_{\nu^*} = 300$~GeV.

\section{Interpretation}

The event yields observed in all decay  channels are in agreement with the corresponding SM expectations and are summarised in table~\ref{tab:nustaryields}. 
The SM predictions are dominated by NC DIS processes for the $e{q}{\bar{q}}$ resonance search and by CC DIS for the $\nu\gamma$ and $\nu{q}{\bar{q}}$ resonance searches. 
The distributions of the invariant mass of the data events are in agreement with those of the expected SM background as shown in figure~\ref{fig:Mass}. 
No data event is observed in channels corresponding to leptonic decays of the $W$ or $Z$ bosons, in agreement with the low SM expectations.

Since no evidence for the production of  excited neutrinos is observed, upper limits on the $\nu^*$ production cross section and on the coupling $f/{\Lambda}$ are derived as a function of the mass of the excited neutrino. 
Limits are presented at the $95$\% confidence level (CL) and are obtained from the mass spectra using a modified frequentist approach which takes statistical and systematic uncertainties into account~\cite{Junk:1999kv}.

\begin{figure}[htbp] 
  \begin{center}
\includegraphics[width=.5\textwidth]{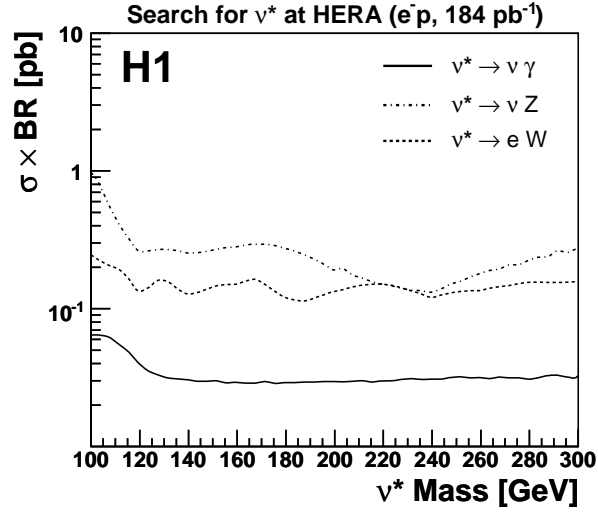}
   \end{center}
  \caption{Upper limits at $95$\% CL on the product of the $\nu^*$ production cross section and decay branching ratio, $\sigma \times$~BR, in the three decay channels as a function of the excited neutrino mass. The different decay channels of the $W$ and $Z$ gauge bosons are combined. Areas above the curves are excluded. }
\label{fig:LimitSBR}  
\end{figure}

\begin{figure}[htbp]
  \begin{center}
\includegraphics[width=.5\textwidth]{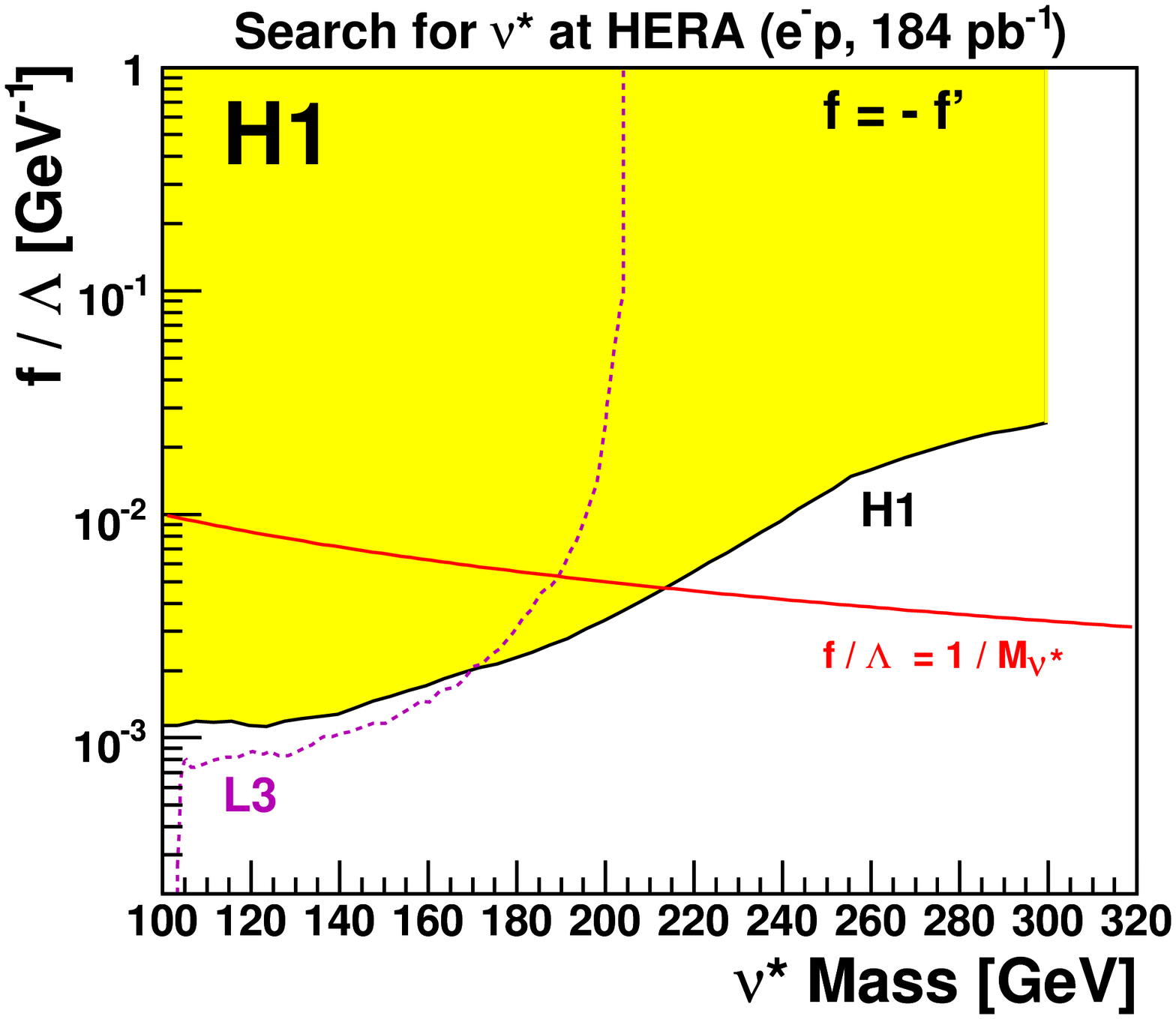}\put(-10,13){{\bf (a)}}
\includegraphics[width=.5\textwidth]{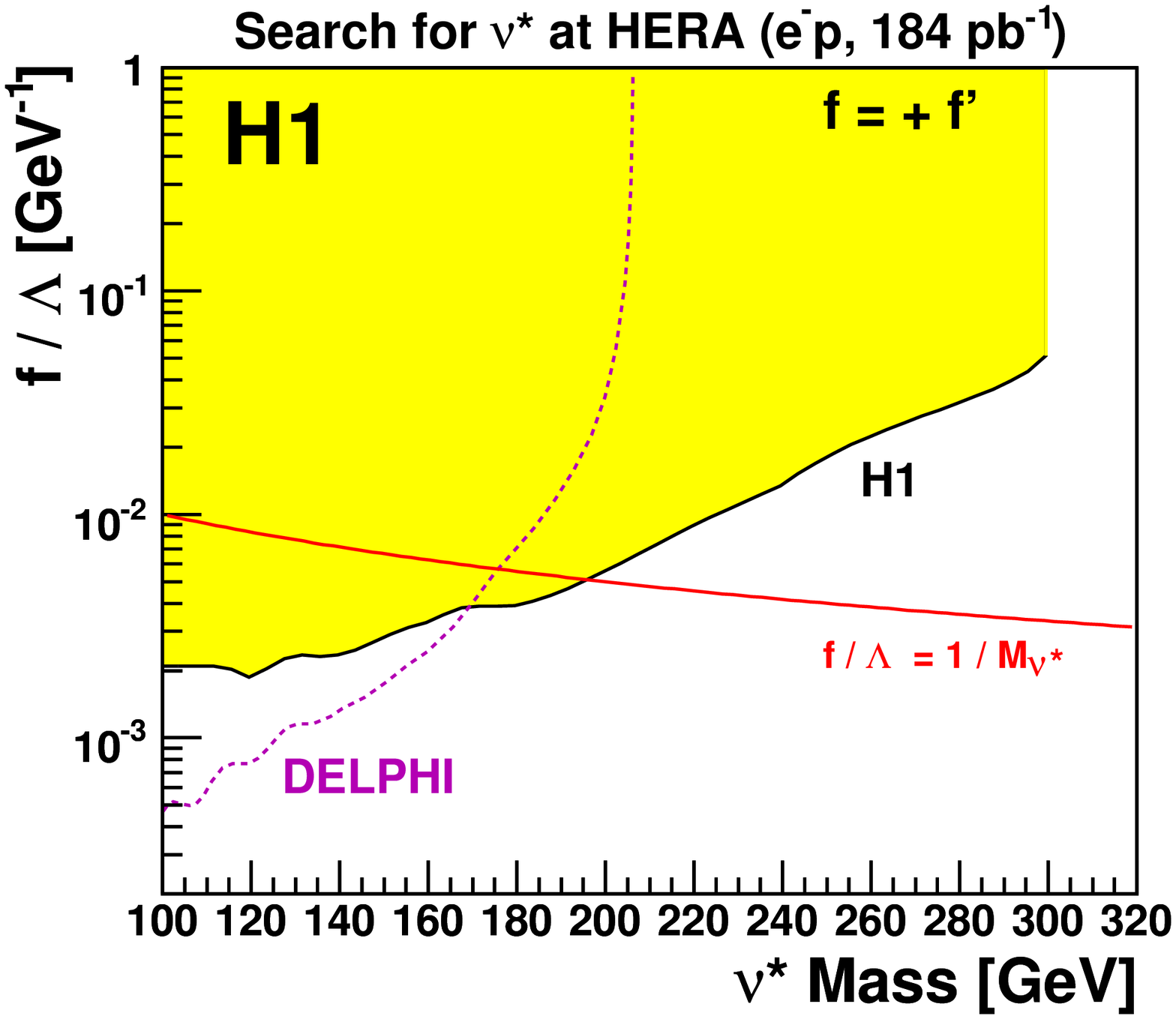}\put(-10,13){{\bf (b)}}
  \end{center}
  \caption{Exclusion limits at $95$\% CL on the coupling $f/\Lambda$ as a function of the mass of the excited neutrino with the assumptions (a) $f = -f'$ and (b) $f = +f'$. 
The excluded domain based on  all H1 $e^-p$ data is represented by the shaded area.
Values of the couplings above the curves are excluded.
The dashed line corresponds to the exclusion limit obtained at LEP by the L3 Collaboration~\protect{\cite{Achard:2003hd}} in (a) and by the DELPHI Collaboration~\protect{\cite{Abdallah:2004yn}} in (b). }
\label{fig:LimitCoupling}  
\end{figure}

Upper limits on the product of the $\nu^*$ production cross section and decay branching ratio are shown in figure~\ref{fig:LimitSBR}. 
The analysed decay channels of the $W$ and $Z$ gauge bosons are combined.
The resulting limits on $f/\Lambda$ after combination of all $\nu^*$ decay channels are displayed as a function of the $\nu^*$ mass in figure~\ref{fig:LimitCoupling}, for the conventional assumptions $f = - f'$ and $f = + f'$. 
Limits are derived for $\nu^*$ masses up to $300$~GeV.
The total fraction of $\nu^*$ decays covered in this analysis is $\sim 92$\% and $\sim 84$\% in the two cases $f = - f'$ and $f = + f'$, respectively.
In the case $f = - f'$, the limit on $f/\Lambda$ is dominated at low mass by the $\nu^* {\rightarrow} \nu \gamma$ channel, while the $\nu^* \rightarrow e W$ channel starts to contribute significantly for masses above $200$~GeV.
Under the assumption $f = + f'$, the limit on $f/\Lambda$ is driven mainly by the $\nu^* {\rightarrow} e W$ channel.
These new results improve significantly the previously published limits by H1~\cite{Adloff:2001me} and ZEUS~\cite{Chekanov:2001xk}.
For comparison, the most stringent limits obtained in $e^+ e^-$ collisions at LEP for the two cases $f = - f'$ and $f = + f'$, determined by L3~\cite{Achard:2003hd} and DELPHI~\cite{Abdallah:2004yn}, respectively, are also shown in figure~\ref{fig:LimitCoupling}.
The H1 measurement provides the most stringent constraints for masses larger than $\sim$~$170$~GeV.
With the assumption $f/\Lambda = 1/M_{\nu^*}$ excited neutrinos with masses up to $213$~GeV ($196$~GeV) are excluded for $f = - f'$ ($f = + f'$).

\begin{figure}[htbp]
  \begin{center}
\includegraphics[width=.5\textwidth]{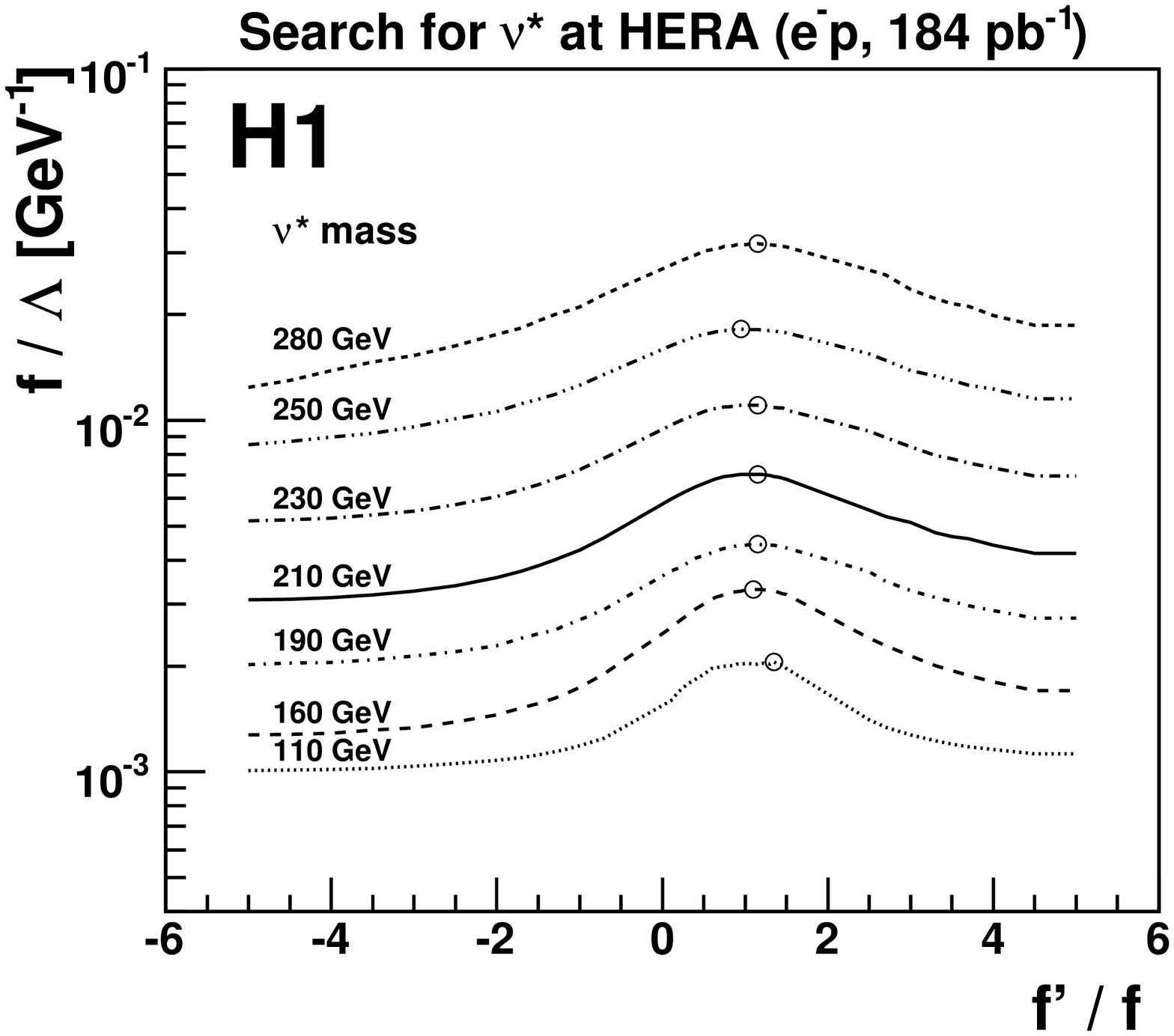}\put(-9,58){{\bf (a)}}
\includegraphics[width=.5\textwidth]{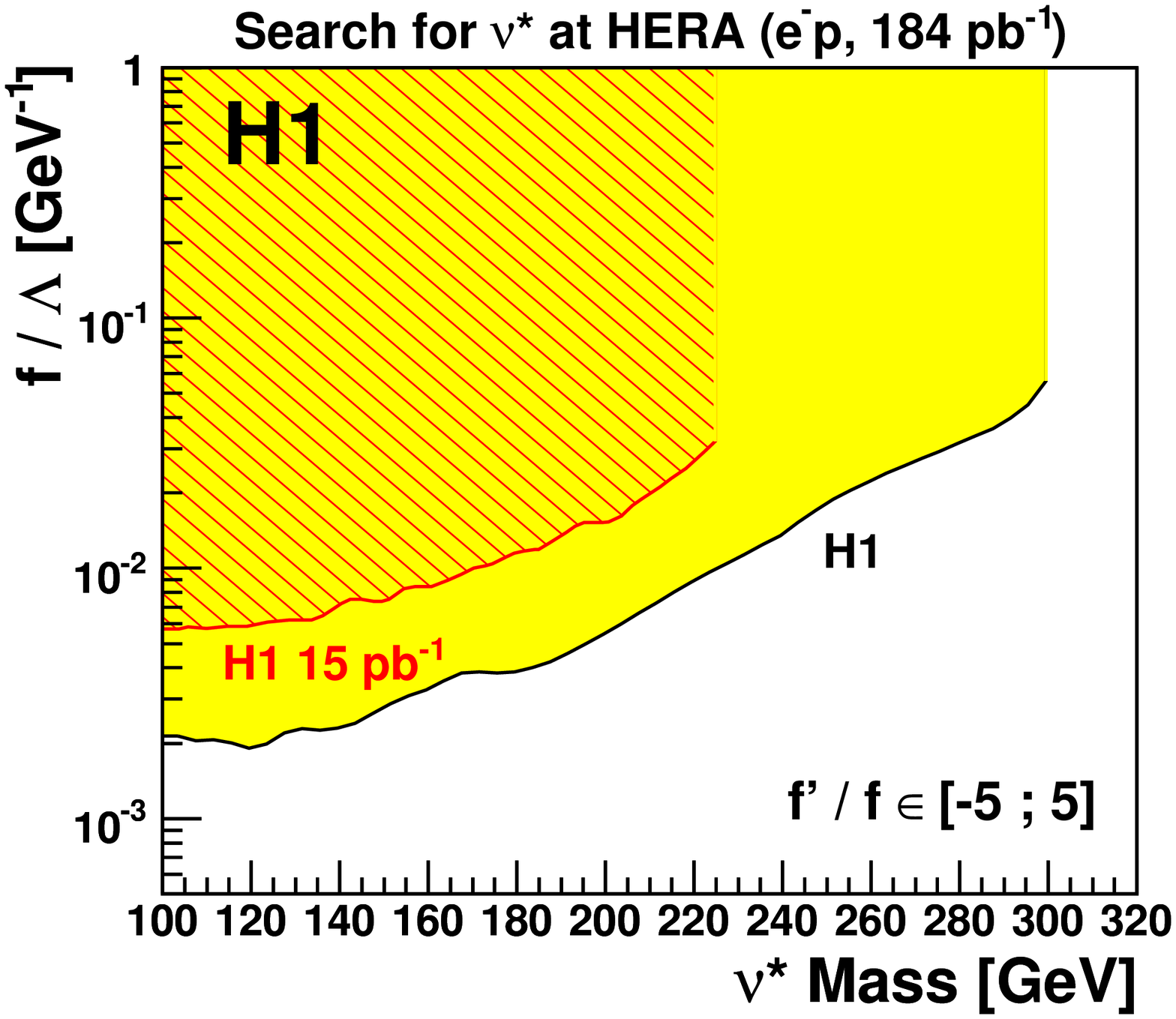}\put(-9,58){{\bf (b)}}
  \end{center}
  \caption{(a) Exclusion limits at $95$\% CL on the coupling $f/\Lambda$ as a function of the ratio $f'/f$.
  Each curve corresponds to a different $\nu^*$ mass. The circle indicates the weakest limit for each mass.
  (b) Exclusion limit at $95$\% CL on the coupling $f/\Lambda$ as a function of the mass of the excited neutrino (shaded area). This limit corresponds to the weakest limit on $f/\Lambda$ for $f'/f$ values in the interval $\left[ -5;+5 \right]$.   
The hatched area corresponds to the exclusion domain obtained by H1 in a previous analysis~\cite{Adloff:2001me}.
The regions above the lines are excluded.}
\label{fig:Limit_fpof}  
\end{figure}

Limits with less model dependence can be derived if arbitrary ratios $f'/f$ are considered.
The dependence of the limits on this ratio for different $\nu^*$ masses is displayed in figure~\ref{fig:Limit_fpof}(a).
Limits which are independent of $f'/f$ are derived for $f'/f \in \left[ -5; 5 \right]$ by choosing in figure~\ref{fig:Limit_fpof}(a) the point with the weakest limit for each mass hypothesis.
The result is shown in figure~\ref{fig:Limit_fpof}(b) and is found to be almost equal to the limit obtained under the assumption $f = + f'$.

\section{Conclusion}

Using the full $e^{-}p$ data sample collected by the H1 experiment at HERA with an integrated luminosity of $184$~pb$^{-1}$ a search for the production of excited neutrinos is performed. 
The excited neutrino decay channels ${\nu}^{*} {\rightarrow} {\nu}{\gamma}$,  ${\nu}^{*} {\rightarrow} {\nu}{Z}$ and ${\nu}^{*} {\rightarrow} {e}{W}$ with subsequent hadronic or leptonic decays of the $W$ and $Z$ bosons are considered and no indication of a $\nu^*$ signal is found.
New limits on the production cross section of excited neutrinos are obtained. 
Previous HERA results are improved by a factor three to four.
Upper limits on the coupling $f/\Lambda$ as a function of the excited neutrino mass are established for specific relations between the couplings ($f = + f'$ and $f = - f'$) and independent of the ratio $f'/f$.
Assuming $f = - f'$ and $f/\Lambda=1/M_{\nu^*}$, excited neutrinos with a mass lower than $213$~GeV are excluded at $95$\% confidence level.
The results presented in this letter greatly extend the previously excluded domain and demonstrate the unique sensitivity of HERA to excited neutrinos with masses beyond the LEP reach.

\section*{Acknowledgements}

We are grateful to the HERA machine group whose outstanding
efforts have made this experiment possible. 
We thank the engineers and technicians for their work in constructing 
and maintaining the H1 detector, our funding agencies for financial 
support, the DESY technical staff for continual assistance and the 
DESY directorate for the hospitality which they extend to the non DESY 
members of the collaboration.
We would like to thank M. Spira for helpful discussions.


\clearpage

\begin{table}[]
\begin{center}
\begin{tabular}{l c c c}

\multicolumn{4}{c}{{\bf Search for \begin{boldmath}$\nu^*$\end{boldmath} at HERA (\begin{boldmath}$e^{-}p$\end{boldmath}, \begin{boldmath}$184$\end{boldmath} pb\begin{boldmath}$^{-1}$\end{boldmath})}}\\
\hline
Channel & Data & SM & Signal Efficiency [\%]\\
\hline
${\nu}^{*} {\rightarrow} {\nu}{\gamma}$ & $7$ & $12.3~{\pm}~3.0$~~ & $50$--$55$\\

${\nu}^{*} {\rightarrow} {e} W {\rightarrow} eq\bar{q}$ & $220$ & $223~{\pm}~47$~~ &  $40$--$65$\\

${\nu}^{*} {\rightarrow} {e} W {\rightarrow} e\nu\mu$  & $0$ & $0.40~{\pm}~0.05$& $35$\\

${\nu}^{*} {\rightarrow} {e} W {\rightarrow} e \nu e$ & $0$ & $0.7~{\pm}~0.1$ & $45$ \\

${\nu}^{*} {\rightarrow} {\nu} Z {\rightarrow} \nu q\bar{q}$ & $89$ & $95~{\pm}~21$ &  $25$--$55$\\

 ${\nu}^{*} {\rightarrow} {\nu} Z {\rightarrow} \nu ee $ & $0$ & $0.19~{\pm}~0.05$ &  $45$\\
\hline
\end{tabular}
\end{center}

\caption{
Observed and predicted event yields for the studied $\nu^*$ decay channels.
  The analysed data sample corresponds to an integrated luminosity of $184$~pb$^{-1}$.
  The error on the SM predictions includes model and experimental systematic errors added in quadrature.
  Typical selection efficiencies for $\nu^*$ masses ranging from $120$ to $260$~GeV are also indicated.
}
\label{tab:nustaryields}
\end{table}

\clearpage

\vspace*{2.5cm}

\end{document}